\documentclass[12pt]{iopart}
\usepackage{graphicx}
\usepackage{epstopdf}
\newcommand{\beq}{\begin{equation}}
\newcommand{\eeq}{\end{equation}}
\newcommand{\bea}{\begin{eqnarray}}
\newcommand{\eea}{\end{eqnarray}}

\usepackage{iopams}
\begin{document}

\title[Strange metals and the AdS/CFT correspondence]{Strange metals and the AdS/CFT correspondence}

\author{Subir Sachdev}

\address{Department of Physics, Harvard University, Cambridge MA
02138, USA}
\ead{sachdev@physics.harvard.edu}
\begin{abstract}
I begin with a review of quantum impurity models in condensed matter physics, 
in which a localized spin degree of freedom
is coupled to an interacting conformal field theory in $d=2$ spatial dimensions. Their properties
are similar to those of supersymmetric generalizations which can be solved by the AdS/CFT
correspondence; the low energy limit of the latter models is described by a AdS$_2$ geometry.
Then I turn to Kondo lattice models, which can be described by a mean-field theory obtained by 
a mapping to a quantum impurity coupled to a self-consistent environment. Such a theory yields a 
`fractionalized Fermi liquid' phase
of conduction electrons coupled to a critical spin liquid state, and is
an attractive mean-field theory of strange metals.
The recent holographic description of strange metals with a AdS$_2 \times R^2$ geometry
is argued to be related to such mean-field solutions of Kondo lattice models.
\end{abstract}

\submitto{Special Issue of JSTAT.\\ Plenary talk at Statphys24, Cairns, Australia, July 2010}
\maketitle

\section{Introduction}
\label{sec:intro}

The `strange metal' is a non-zero temperature phase of electrons in solids which appears to be a common to
most `correlated electron' compounds. These are compounds with transition-metal or rare-earth
elements with a crystal structure which often promotes electronic transport primarily within a single two-dimensional
layer of atoms. The strange metal is typically found at temperatures ($T$) above low temperature phases which
display antiferromagnetism and superconductivity. It is defined as `strange' because the temperature and frequency
dependence of many observables deviate strongly from those expected from conventional Fermi liquid theory:
most famously, the resistance increases linearly with $T$, over a wide range of $T$ values.  
While the cuprate high temperature superconductors are the most prominent compounds displaying strange metal
behavior \cite{cooper}, similar regimes are also found in 
ruthenium oxides \cite{grigera}, iron pnictides \cite{kasahara}, organic metals \cite{doiron}, and
heavy-fermion compounds \cite{gegenwart}.

The AdS/CFT correspondence was originally discovered as a tool for describing strongly-coupled
gauge theories \cite{MAGOO}. In $D=4$ spacetime dimensions, Yang Mills gauge theory with 
$\mathcal{N}=4$ supersymmetry (abbreviated as SYM4) 
is characterized by a single dimensionless coupling constant which remains
invariant under the renormalization group (RG). Consequently, the theory is conformally invariant for 
all values of the coupling, and can be viewed as the simplest interacting conformal field theory
in $D=4$ spacetime dimensions: a CFT4. With a SU($M$) gauge group, this theory was argued to be
equivalent to a string theory on an AdS$_5 \times S^5$ background, where AdS
is anti-de Sitter, a symmetric space with constant negative curvature. More usefully, in the $M\rightarrow \infty$ limit
for the low energy physics, the string theory can be approximated simply by classical Einstein gravity
on AdS$_5$. Thus, remarkably, correlations of gauge theories in $D=4$ can be related to properties
of the Einstein gravity in a negatively curved space in $D=5$. The latter space has an emergent dimension,
which can be interpreted as a RG energy scale.

CFTs also arise in condensed matter physics in many different contexts. In Section~\ref{sec:qafm}, we will briefly
review the CFT3s arising near the quantum critical points of certain quantum antiferromagnets in $d=2$ spatial
dimensions. It was argued in Ref.~\cite{m2cft} that such CFT3s could also be usefully analyzed via the AdS/CFT correspondence. In $D=3$, a supersymmetric analog of the CFT3s in Section~\ref{sec:qafm} is Yang-Mills
theory with $\mathcal{N}=8$ supersymmetry and a SU($M$) gauge group (SYM3), and 
this maps in the $M\rightarrow \infty$ limit
to gravity on AdS$_4$. Unlike the $D=4$ case, the coupling constant of SYM3
flows generically to a strong-coupling fixed point, and so there is no free coupling and the low energy physics
is conformal. Thus it is always `quantum critical', and it is the first solvable strongly-interacting quantum critical theory
in $D=3$.
The condensed matter CFT3s of Section~\ref{sec:qafm} also have strongly interacting quantum critical points,
many of whose properties have resisted accurate solution by other available methods. Even though they are not realized
as the large $M$ limit of a non-Abelian gauge theory, it was argued \cite{m2cft} that they could be modeled via
the AdS/CFT correspondence. In this context, it should be kept in mind that the operator content of the 
strongly-coupled fixed point of SYM3 is far removed from the non-Abelian gauge fields
in terms of which it was written down at high energy scales; thus its quantum critical physics is no more the physics
of non-Abelian gauge fields than that of the models of Section~\ref{sec:qafm}. Indeed, it is not unreasonable to view SYM3
at small $M$ as a supersymmetric generalization of the famous classical $D=3$ Ising model at its critical point.
Such applications of the AdS/CFT correspondence to CFTs in  condensed matter have been reviewed by the author
in other articles \cite{lt,milos}, 
and so will not be explored here. In a recent paper \cite{myers}, we have argued that the AdS/CFT results can
yield useful information for the frequency and temperature dependence of transport co-efficients which 
have been studied in experiments in a variety of condensed matter systems.

The CFTs discussed so far, and those in Section~\ref{sec:qafm}, are states of quantum matter which are effectively
at {\em zero density}. Their low energy spectrum is relativistic and particle-hole symmetric, similar to that
found in {\em e.g.\/} pure, undoped graphene. To access states analogous to the strange metal, we have
to study quantum matter at non-zero density. We do this by turning on a chemical potential, $\mu$, which couples to a 
globally conserved charge of the CFT; we choose $\mu=0$ to correspond to the zero density state.
As long as $|\mu| \ll T$, the AdS/CFT correspondence can be extended straightforwardly, and many new results
for quantum-critical transport in condensed matter have been obtained by this method \cite{nernst}. These have also been
reviewed elsewhere \cite{lt,milos}, and will not be discussed further here.

We turn, finally, to states of non-zero density quantum matter at low temperature, which have $T \ll |\mu|$.
Here, even for the supersymmetric CFTs at large $M$, the application of the AdS/CFT correspondence
is not immediate. The CFT has scalar fields with exactly flat directions in their potential, and 
when placed at a non-zero $\mu$ at $T=0$ these flat directions appear to lead to an instability of the theory.
Nevertheless,  one can presume that the strongly coupled CFT
at non-zero $\mu$ continues to have a gravitational description in the AdS language. In the absence of a precise
derivation of the theory on AdS, the spirit of effective field theory can be used to postulate a phenomenological
action on AdS, which then predicts interesting new physics in the doped-CFT at low $T$. Just such a strategy
has been used in a large number of recent studies. Two broad classes of states have been obtained by this method.
States in one class display the condensation of a charged \cite{gubser,hhh,shiraz,double} or a neutral \cite{sandip,tavanfar} scalar field
(or both \cite{siliu});
we will not study this class here.
States in the second class have no broken symmetries and display evidence of metallic behavior
and Fermi surfaces \cite{Lee,Cubrovic,Liu,Faulkner,denef,tong,FP,mfl,larsen,elias,ssffl,physics}. 
Interestingly, the Fermi surface quasiparticles have non-Fermi liquid damping,
and the resistance can have a linear dependence on $T$ for a particular value of an effective parameter
{\em i.e.\/} the AdS theory at non-zero $\mu$ leads to a holographic description of the strange metal.

However, many key properties of the holographic strange metal so obtained are quite mysterious from a 
condensed-matter perspective. For a CFT$D$,
the AdS$_{D+1}$ theory at non-zero $\mu$ factorizes \cite{Faulkner} at low energies to a AdS$_2 \times R^{D-1}$ geometry,
and this factorization is a key determinant in the unusual properties of the Fermi surface. One immediate consequence
is that the singular features in the quasiparticle energy depend primarily on the frequency $\omega$, and this
quantum criticality is described by the AdS$_2$ geometry alone: thus crucial information on the spatial
dependence of the strange metal quantum fluctuations appears to be missing. Related to this is the troubling
presence of a non-zero entropy density which survives the $T \rightarrow 0$ limit, thus apparently violating
the third law of thermodynamics. Finally, the underlying field content leading to this holographic metal phase
is not clear.

This paper will review and give additional perspective on a recent proposal \cite{ssffl} connecting Kondo lattice models
to the holographic strange metal. We will do this here by adding matter degrees of freedom to the zero density CFT 
`one-at-a-time'. As noted above, we begin in Section~\ref{sec:qafm} by describing how CFTs arise
in two-dimensional quantum antiferromagnets. In Section~\ref{sec:qimp} will add a single spin degree of freedom to the CFT: models of this
type can be reliably addressed both by traditional condensed matter methods, and in supersymmetric cases
by the AdS/CFT correspondence. We will find a close correspondence in the physical properties in the two cases,
including an emergence of AdS$_2$ in the gravitational description. Thus in the single impurity context,
a satisfactory physical interpretation of the AdS$_2$ geometry will be obtained.

Then we will turn to Kondo lattice models and their holographic interpretation in Section~\ref{sec:lattice}.
In the limit of large dimension, or long-range exchange interactions, such models can be solved
by a mapping to a quantum impurity model coupled to a self-consistent environment. 
The mean-field solution describes a `fractionalized Fermi liquid' (FFL or FL*), with a Fermi surface of conduction
electrons coupled to a critical spin liquid.
We will argue that
such a solution 
has a close correspondence to the holographic theory with the AdS$_2 \times R^{D-1}$ geometry.
This description of the holographic theory leads to simple interpretations of its
physical properties. For the Kondo lattice, the mean-field theory of the critical spin liquid does not
have collective gauge excitations which are generically expected to be present in realistic spin liquid
with finite-range interactions. The holographic theory reaches a similar mean-field but without
infinite range interactions: thus it is likely to be amenable to systematic improvements. We hope such
improvements will eventually lead to a description of realistic spin liquids and FL* states.

\section{Quantum antiferromagnets and CFT3s}
\label{sec:qafm}

We begin with a brief review of the connection between lattice quantum antiferromagnets
in $d=2$ and CFT3s, also reviewed in \cite{milos}. The lattice antiferromagnets are described 
by the Heisenberg exchange Hamiltonian
\begin{equation}
H_J = \sum_{i<j} J_{ij} \hat{S}^a_i \, \hat{S}^a_j + \ldots
\label{eq:ssHJ}
\end{equation}
where $J_{ij} > 0$ is the antiferromagnetic exchange interaction and $\hat{S}^a_i$ ($a=x,y,z$) are $S=1/2$
spin operators on the sites, $i$, of a regular lattice: thus they obey the commutation relations
\begin{equation}
[ \hat{S}^a_i, \hat{S}^b_j ] = i \epsilon_{abc} \delta_{ij} \hat{S}^c_i
\label{commute}
\end{equation}
and $\sum_a (\hat{S}^a_i )^2 = 3/4$ for each $i$.

Different phases, quantum phase transitions and low energy field theories are obtained depending upon 
whether there are an even or an odd number
of $S=1/2$ spins per unit cell of the lattice. 

\subsection{Even number of $S=1/2$ spins per unit cell}
\label{sec:even}

\begin{figure}
\centering
 \includegraphics[width=5in]{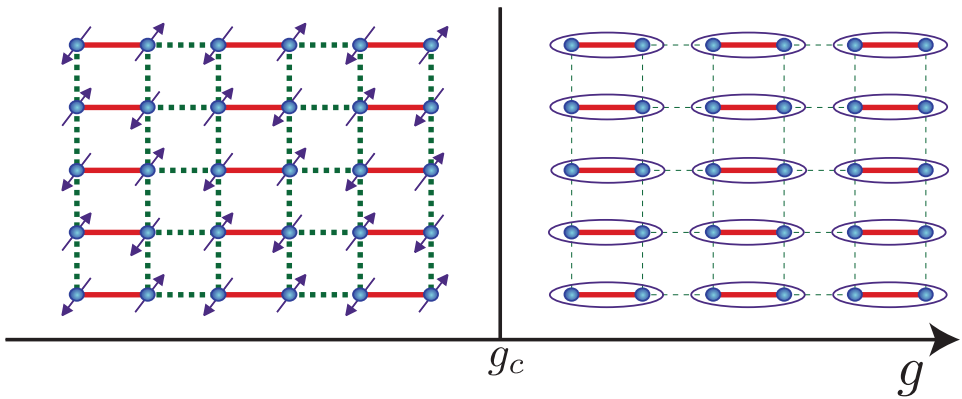}
 \caption{The dimer antiferromagnet. The full red lines represent an exchange interaction $J$, while the dashed green lines have exchange $J/g$. The ellispes represent a singlet valence
 bond of spins $(|\uparrow \downarrow \rangle - | \downarrow \uparrow \rangle )/\sqrt{2}$.}
\label{fig:ssdimer}
\end{figure}

Let us consider first the simpler case of an even number of $S=1/2$ spins per unit cell.
The canonical model is the dimer antiferromagnet, illustrated in Fig.~\ref{fig:ssdimer}.
The $S=1/2$ spins reside on the sites of a square lattice, and have nearest neighbor exchange equal
to either $J$ or $J/g$. Here $g \geq 1$ is a tuning parameter which induces a quantum phase
transition in the ground state of this model. 
At $g = 1$, the model has full square lattice symmetry,
and this case is known to have a N\'eel ground state which breaks spin rotation symmetry. This state has a
checkerboard polarization of the spins, just as found in the classical ground state, and as illustrated 
on the left side of Fig.~\ref{fig:ssdimer}. It can be characterized by a vector order parameter $\varphi^a$
which measures the staggered spin polarization
\begin{equation}
\varphi^a = \eta_i S^a_i
\end{equation}
where $\eta_i=\pm 1$ on the two sublattices of the square lattice. In the N\'eel state we have $\langle \varphi^a \rangle \neq  0$, and we expect that the low energy excitations can be described by long wavelength fluctuations 
of a field $\varphi^a (r, \tau)$ over
space, $r$, and imaginary time $\tau$.
On the other hand, for $g \gg 1$ it is evident from Fig.~\ref{fig:ssdimer} that the ground state preserves 
all symmetries of the Hamiltonian: it has total spin $S=0$ and can be considered to be a product of nearest
neighbor singlet valence bonds on the $J$ links. The simplicity of this large $g$ ground state relies crucially
on the `dimerized' structure of the Hamiltonian; the fact that there are an even number of $S=1/2$ spins
per unit cell.

It is clear that the $g=1$ and $g \gg 1$ states are qualitatively distinct, and so there must be a quantum phase 
transition at a critical $g=g_c$. We can deduce the quantum field theory for this phase transition
by using conventional Landau-Ginzburg arguments: we use the order parameter $\varphi^a$, and write
down the simplest continuum action consistent with the symmetries of the Hamiltonian; this leads to the
partition function
\begin{eqnarray}
\mathcal{Z} &=& \int \mathcal{D} \varphi^a (r, \tau) \exp \left( - \int d^2 r d \tau \, \mathcal{L}_\varphi \right) \nonumber \\
\mathcal{L}_\varphi &= & \frac{1}{2} \left[ (\partial_\tau \varphi^a )^2  + v^2 ( \nabla \varphi^a )^2 + s ( \varphi^a)^2 \right]
+ \frac{u}{4} \left[ (\varphi^a)^2 \right]^2 \label{zphi}
\end{eqnarray}
The transition is now tuned by varying $s \sim (g - g_c)$. Notice that this model 
is identical to the Landau-Ginzburg theory for the thermal phase transition in a $d+1$ dimensional ferromagnet,
because time appears as just another dimension.
From this we conclude that the quantum phase transition is described by the famous
Wilson-Fisher fixed point of Eq.~(\ref{zphi}). This was originally discovered by an analysis of the theory
in $D=4-\epsilon$ spacetime dimensions, using $\epsilon$ as an expansion parameter. Since then, extensive
numerical and analytical studies have shown that the fixed point is present also in $D=3$, where it describes
a non-trivial CFT3. For the dimer antiferromagnet, very convincing evidence that the quantum criticality
is described by the Wilson-Fisher CFT3 is presented in \cite{ssjanke}.

In experiments, the best studied realization of the dimer antiferromagnet is TlCuCl$_3$. In this crystal, the dimers are coupled
in all three spatial dimensions, and the transition from the dimerized state to the N\'eel state can be induced by application of pressure.
Neutron scattering experiments by Ruegg and collaborators \cite{ssruegg} have 
clearly observed the transformation in the excitation spectrum across the transition,
and these observations are in good quantitative agreement with theory\cite{sssolvay}.

\subsection{Odd number of $S=1/2$ spins per unit cell}
\label{sec:odd}

\begin{figure}
\centering
 \includegraphics[width=5in]{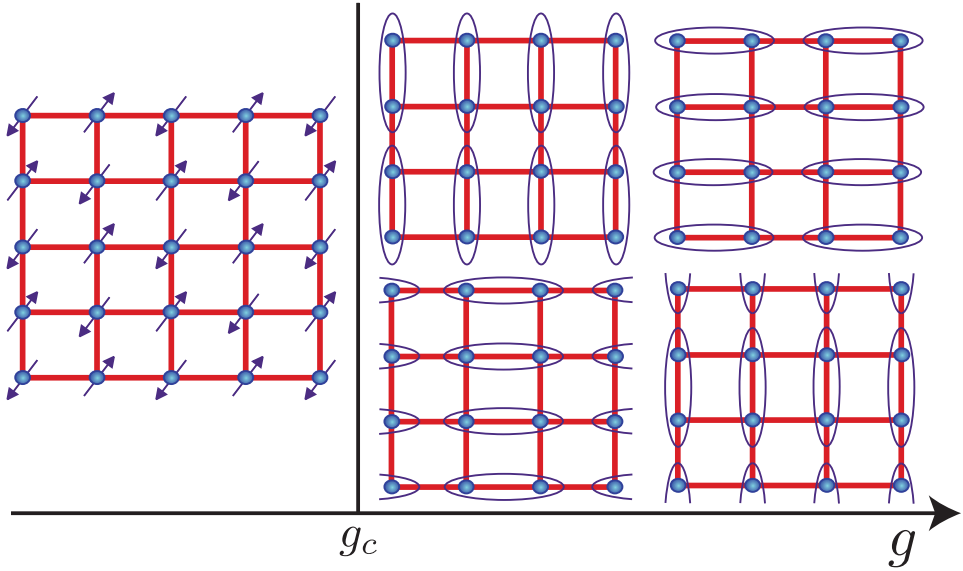}
 \caption{A frustrated square lattice antiferromagnet on the square lattice, with the Hamiltonian preserving the full 
 space group symmetry of the square lattice. The valence bond solid (VBS) state for $g>g_c$
 is four-fold degenerate, depending upon the crystallization pattern of the singlet valence bonds.}
\label{fig:vbs}
\end{figure}

For this case, we can work with the Hamiltonian in Eq.~(\ref{eq:ssHJ}), but with the $J_{ij}$ respecting the full space group
symmetry of the square lattice. Thus the nearest neighbor $J_{ij}$ must all be equal to each other, unlike
the dimer antiferromagnet above. With no further range interactions, the ground state has N\'eel order,
as we discussed in Section~\ref{sec:even}. A variety of routes have been investigated \cite{sssolvay} to continuously
destroy the N\'eel order, and we generically represent them as tuning a coupling $g$ in Fig.~\ref{fig:vbs}.
The phase so-reached was argued \cite{ssrsl} to have valence bond solid order (VBS).
The VBS state is superficially similar to the dimer singlet state in the right panel of Fig.~\ref{fig:ssdimer}:
the spins primarily form valence bonds with near-neighbor sites. However, because of the square lattice symmetry of the Hamiltonian, a columnar arrangement of the valence bonds as in Fig.~\ref{fig:ssdimer}, breaks the square lattice rotation
symmetry; there are 4 equivalent columnar states, with the valence bond columns running along different directions. 
More generally, a VBS state is a spin singlet state, with a non-zero degeneracy due to a spontaneously broken lattice
symmetry. 
A VBS state has been observed in the organic antiferromagnet EtMe$_3$P[Pd(dmit)$_2$]$_2$ \cite{sskato1,sskato2}.  

A direct transition at $g=g_c$ between the N\'eel and VBS states involves two distinct broken symmetries:
spin rotation symmetry, which is broken only in the N\'eel state for $g<g_c$, and a lattice rotation symmetry,
which is broken only
in the VBS state for $g>g_c$. The rules of Landau-Ginzburg-Wilson theory imply that there can be no generic second-order
transition between such states.
It has been argued that a second-order N\'eel-VBS transition can indeed occur \cite{sssenthil}, but the critical theory is not expressed
directly in terms of either order parameter. It involves a fractionalized bosonic spinor $z^\alpha$ ($\alpha = \uparrow,
\downarrow$), and an emergent gauge field $A_\mu$. 
The key step is to express the vector field $\varphi^a$ in terms of $z^\alpha$ by
\begin{equation}
\varphi^a = z_\alpha^\ast (\sigma^a)^{\alpha}_{\beta} z^\beta
\label{eq:ssPhiz}
\end{equation}
where ${\sigma}^a$ are the $2\times2$ Pauli matrices. Note that this mapping from $\varphi^a$ to $z^\alpha$
is redundant. We can make a spacetime-dependent change in the phase of the $z^\alpha$ by the field $\theta(r,\tau)$
\begin{equation}
z^\alpha \rightarrow e^{i \theta} z^\alpha
\label{eq:ssgauge}
\end{equation}
and leave $\varphi^a$ unchanged. All physical properties must therefore also be invariant under Eq.~(\ref{eq:ssgauge}),
and so the quantum field theory for $z^\alpha$ has a U(1) gauge invariance, much like that found in quantum electrodynamics.
The effective action for the $z^\alpha$ therefore requires introduction of an `emergent' 
U(1) gauge field $A_\mu$ (where $\mu = x, \tau$ is a 
three-component spacetime index). The field $A_\mu$ is unrelated the electromagnetic field, but is an internal
field which conveniently describes the couplings between the spin excitations of the antiferromagnet.  
As for Eq.~(\ref{zphi}),
we can write down the quantum field theory for $z^\alpha$ and $A_\mu$ by the constraints of symmetry and gauge invariance,
which now yields
\begin{eqnarray}
\mathcal{Z} &=& \int \mathcal{D} z^\alpha (r, \tau) \mathcal{D} A_\mu (r, \tau) \exp \left( - \int d^2 r d \tau \,
\mathcal{L}_z \right) \nonumber \\
\mathcal{L}_z &=&  
|(\partial_\mu -
i A_{\mu}) z^\alpha |^2 + s |z^\alpha |^2  + u (|z^\alpha |^2)^2 + \frac{1}{2w^2}
(\epsilon_{\mu\nu\lambda}
\partial_\nu A_\lambda )^2  \label{zz}
\end{eqnarray}
For brevity, we have now used a ``relativistically'' invariant notation, and scaled away the spin-wave velocity $v$; the values
of the couplings $s,u,w$ are different from, but related to, those in Eq.~(\ref{zphi}). The Maxwell action for $A_\mu$ is generated from 
short distance $z^\alpha$ fluctuations, and it makes $A_\mu$ a dynamical field.
This theory has a `Higgs' phase where $z^\alpha$ condenses like the Higgs boson: this we can identify
as the N\'eel state. The ordinary Coulomb phase with $z^\alpha$ gapped appears as a `spin liquid' state with a collective gapless, spinless excitation associated with the 
$A_\mu$ photon. Non-perturbative effects \cite{ssrsl} associated with the monopoles in $A_\mu$ (not discussed here), show that this spin liquid is
ultimately unstable to the appearance of VBS order. 

An interesting question now is whether the transition between the N\'eel and VBS states as described by
(\ref{zz}) is a CFT3. The existence of a CFT3 fixed point has been established order-by-order in the $1/N$ expansion,
where the spinor index $\alpha = 1 \ldots N$. However, the issue remains unsettled for $N=2$ \cite{sssandvik,alet}.

For our purposes here, the CFT3s described by Eqs.~(\ref{zphi}) and (\ref{zz}) are non-supersymmetric analogs
of the CFT3 realized by SYM3. Insights gained from the AdS/CFT correspondence are described elsewhere \cite{milos,myers}.

\section{Quantum impurity in a CFT}
\label{sec:qimp}

As we discussed in Section~\ref{sec:intro}, we will move away from the zero density CFTs of Section~\ref{sec:qimp}
by a adding a single defect localized in space. This will eventually allow us to address the non-zero density case in 
the following section.

\begin{figure}
\centering
 \includegraphics[width=4in]{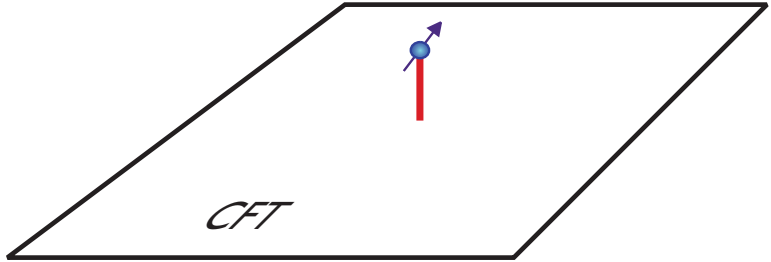}
 \caption{A quantum spin coupled via an exchange interaction to a CFT in 2+1 dimensions.}
\label{fig:qimp}
\end{figure}

For the quantum antiferromagnets of Section~\ref{sec:qimp}, the simplest interesting defect is a single spin $\hat{S}^a$
coupled to the antiferromagnet by an exchange coupling, $J$, as shown in Fig.~\ref{fig:qimp}.

More generally, the configuration of Fig.~\ref{fig:qimp} belongs to a wide class of `Kondo' problems. 
Usually, the bulk CFT is rather simple: it is a free electron system whose Fermi surface excitations form
an infinite set of CFT2s of free chiral fermions in 1+1 spacetime dimensions; furthermore only a single CFT2
of the free fermions is coupled to the impurity. In such cases, all the quantum correlation effects
arise solely in the vicinity of the impurity. In the simplest and most common 
case of quantum spin coupled to a Fermi surface, there
is no interesting quantum-criticality in the low energy limit. There are only two fixed points of the RG, $J=0$
and $J=\infty$: at the unstable $J=0$ fixed point, the impurity decouples from the bulk CFT, while at the 
stable $J=\infty$ fixed point
there are only innocuous potential scattering perturbation \cite{nozieres1}. 
However, the situation becomes far more non-trivial if the bulk fermions acquire an additional `channel' or `flavor'
index; then, under suitable conditions, a non-trivial stable fixed point is obtained at an intermediate $J=J^\ast$ \cite{nozieres2}.
It is the analog of this `multi-channel Kondo fixed point' \cite{affleck} which we will explore in this section in the more
general setting of bulk CFTs with interactions in $D>2$ spacetime dimensions.

As a first example, let us couple the impurity spin $\hat{S}^a$ to the class of dimer antiferromagnets described by the
field theory (\ref{zphi}). We can represent the impurity spin by a coherent state path integral over the fluctuations
of a unit vector $n^a (\tau)$; then the partition function of the quantum impurity problem becomes
\begin{eqnarray}
\mathcal{Z} &=& \int \mathcal{D} \varphi^a (r, \tau) \mathcal{D} n^a (\tau) \delta( [n^a (\tau)]^2 - 1)
\exp \left( - \int d \tau \, \mathcal{L}_{\rm imp} - \int d^2 r d \tau \, \mathcal{L}_\varphi \right) \nonumber \\
 \mathcal{L}_{\rm imp} &=& \frac{i}{2} \mathcal{A}^a  \frac{dn^a}{d \tau} +  J n^a (\tau) \varphi^a (0, \tau)
 \label{zphin}
\end{eqnarray}
Here the quantum spin commutation relations in Eq.~(\ref{commute}) for the impurity spin are implemented
by the Berry phase term in $\mathcal{L}_{\rm imp}$ where $\mathcal{A}^a$ is any function of $n^a (\tau)$ obeying
$\epsilon^{abc} ( \partial \mathcal{A}^b / \partial n^c ) = n^a$. Equivalently, we may represent the impurity spin
by a `slave' fermion $\chi^\alpha$, and then the partition function in Eq.~(\ref{zphin}) can be written as
\begin{eqnarray}
\mathcal{Z} &=& \int \mathcal{D} \varphi^a (r, \tau) \mathcal{D} \chi^\alpha (\tau) 
\exp \left( - \int d \tau \, \mathcal{L}_{\rm imp} - \int d^2 r d \tau \, \mathcal{L}_\varphi \right) \nonumber \\
 \mathcal{L}_{\rm imp} &=& \chi_\alpha^\dagger \frac{\partial \chi^\alpha}{\partial \tau}  +  J \, \chi_\alpha^\dagger \left[ (\sigma^a)^{\alpha}_{\beta} \varphi^a (0, \tau) \right] \chi^\beta \, .
 \label{zphichi}
\end{eqnarray}
Actually, the partition function in Eq.~(\ref{zphichi}) has a conserved fermion number $n_\chi = \chi^\dagger_{\alpha} \chi^\alpha$,
and so splits into different sectors labeled by the possible values of $n_\chi = 0,1,2$. The mapping to Eq.~(\ref{zphin}) requires
restriction to the sector with $n_\chi = 1$; this constraint can be implemented by a Lagrange multiplier, which we
have not written out explicitly.

The quantum impurity problem defined by Eq.~(\ref{zphin}) or (\ref{zphichi})
is amenable to a RG analysis using an expansion in $\epsilon = 4 -D$. An extensive theoretical study has been carried out by
this method \cite{eps1,eps2,eps3}, and we now summarize the main results. 
When the bulk theory is at the Wilson-Fisher CFT3 fixed point, the impurity
coupling $J$ flows to a stable fixed point at some $J=J^\ast$. Some of the characteristics of this fixed point are:
\begin{itemize}
\item The correlations of the impurity fermion $\chi^\alpha$ and the impurity spin $\hat{S}^a =(1/2) \chi_\alpha^\dagger (\sigma^a)^{\alpha}_{\beta} \chi^\beta$ decay with a power-law in time, wtih non-trivial `impurity' exponents which can be computed order-by-order
in $\epsilon$.
\item The impurity response to a uniform external field is characterized by an impurity susceptiblity which
has a Curie form $\chi_{\rm imp} = \mathcal{C}/T$, where $\mathcal{C}$ is a non-trivial universal number which can be computed
in the $\epsilon$ expansion. This response is that of an `irrational' free spin, because $\mathcal{C} \neq S(S+1)/3$, with $2S$ an integer.
\item There is a finite ground state entropy, $S_{\rm imp}$, at $T=0$. This entropy is also `irrational' because $S_{\rm imp}
\neq k_B \ln (\mbox{an integer})$.
\end{itemize}
Extensive numerical studies \cite{test1,test2,test3,test4} 
have also been carried out for the above quantum impurity problem, and the results so far
are in good agreement with the theoretical expectations.

We also mention here an alternative strong-coupling formulation \cite{nls1} of the above quantum impurity coupled to the dimer
antiferromagnet described by $\mathcal{L}_\varphi$. This alternative formulation turns out to be suitable for a determination
of the universal critical properties in an expansion in $\varepsilon = D-2$. For this formulation we turn from the `soft-spin'
formulation of the bulk critical theory in Eq.~(\ref{zphi}), to a `hard-spin' formulation in terms of a unit-length field 
$n^a (r, \tau)$. Then, as is well known, the bulk Wilson-Fisher fixed point is accessed by a $D=2+\varepsilon$ expansion of the 
O(3) non-linear $\sigma$-model. For the quantum impurity physics, it has been argued that the fixed length limit requires
that we send the impurity-bulk coupling $J$ to infinity. In other words, the orientation of the impurity spin $\hat{S}^a$ is fixed
to be parallel to that of the bulk field $n^a (r, \tau)$ at $r=0$. Finally, because the impurity spin orientation is fixed, we only need
a spinless fermion $\chi$ to account for the presence/absence of the impurity. With this reasoning, we obtain the 
partition function
\begin{eqnarray}
\mathcal{Z} &=& \int \mathcal{D} n^a (r, \tau) \mathcal{D} \chi (\tau) \delta( [n^a (r, \tau)]^2 - 1 )
\exp \left( - \int d \tau \, \mathcal{L}_{\rm imp} - \int d^2 r d \tau \, \mathcal{L}_n \right) \nonumber \\
\mathcal{L}_n &=& \frac{1}{2g} \left[ \left( \frac{\partial n^a}{\partial \tau} \right)^2 + c^2 (\nabla n^a )^2 \right] \nonumber \\
 \mathcal{L}_{\rm imp} &=& \chi^\dagger \left( \frac{\partial}{\partial \tau} - i A_\tau \right) \chi \quad , \quad
 A_\tau \equiv \frac{1}{2} \mathcal{A}^a \frac{d n^a (0, \tau)}{d \tau}
 \label{znchi}
\end{eqnarray}
where $\mathcal{A}^a$ is now a function of $n^a (0, \tau)$, and the gauge potential $A_\tau$ is the pullback
of $\mathcal{A}^a$ from $S^2$. 
The conserved fermion number can now only take the values $n_\chi = 0,1$. Here the restriction to $n_\chi =1$ is trivially implemented because there is only a single fermion state without
a fermion spin index:
we simply omit $\chi$ from the functional integral, while including the Wilson line source term $\exp (i \int d \tau A_\tau)$.
The claim \cite{nls1} is that the $\varepsilon=D-2$ expansion of the partition
function in Eq.~(\ref{znchi}) describes the same universal fixed point as the $\epsilon = 4-D$ expansion of the partition
function in Eq.~(\ref{zphichi}). Notice that the theory (\ref{znchi}) has only a single coupling constant $g$, and this reaches
the same fixed point as in the bulk theory. The impurities properties are nevertheless non-trivial and universal, and are entirely
a consequence of the Berry phase of the impurity. The theoretical results from this formulation have been successfully compared
to numerical studies away from the bulk critical point, within the ordered N\'eel phase \cite{nls2,nls3}.

Next, let us turn to the bulk CFT3 associated with an odd-number of $S=1/2$ spins per unit cell, the CP$^{N-1}$ model
in Eq.~(\ref{zz}). Now, there are $S=1/2$ excitations $z^\alpha$ in the bulk, and so the impurity spin can be more
efficiently screened by the environment. To the $z^\alpha$, the impurity spin appears as a static external U(1) gauge charge, and so
the bulk+impurity theory takes a form similar to Eq.~(\ref{znchi}). We have \cite{dc1}:
\begin{eqnarray}
\mathcal{Z} &=& \int \mathcal{D} z^\alpha (r, \tau) \mathcal{D} A_\mu (r, \tau) \mathcal{D} \chi (\tau) 
\exp \left( - \int d \tau \, \mathcal{L}_{\rm imp} - \int d^2 r d \tau \, \mathcal{L}_z \right) \nonumber \\
 \mathcal{L}_{\rm imp} &=& \chi^\dagger \left( \frac{\partial}{\partial \tau} - i A_\tau (0, \tau) \right) \chi 
  \label{zzchi}
\end{eqnarray}
Like Eq.~(\ref{znchi}), there is no coupling constant associated with the impurity theory, and so the impurity responses
are naturally universal. It should be emphasized that the theory in Eq.~(\ref{zzchi}) is {\em different\/} from those in Eqs.~(\ref{zphichi})
and (\ref{znchi}): the bulk CFTs correspond to the two cases in Section~\ref{sec:qafm}, and so the impurity dynamics
is also distinct. The properties of theory in Eq.~(\ref{zzchi}) have been studied in some detail using the $1/N$ expansion \cite{dc1,dc2,dc3,dc4},
and there have also been recent numerical studies of this case \cite{alet,dc5,alet}.

Finally, let us turn to supersymmetric gauge theories, and consider a quantum impurity problem 
associated with a bulk CFT of SYM4 with the SU($M$) gauge group. 
Such a problem was considered recently by Kachru, Karch and Yaida \cite{kky1,kky2}. 
Their impurity was represented by a localized fermion $\chi^b$ with $b=1 \dots M$ a SU($M$) color index. 
The action for their field theory was
\begin{eqnarray}
\mathcal{S} &=& \int d^3 r d \tau \, \mathcal{L}_{\rm SYM} + \int d\tau \, \mathcal{L}_{\rm imp} \nonumber \\
\mathcal{L}_{\rm imp} &=& \chi_b^{\dagger} \frac{\partial \chi^b}{\partial \tau} 
+ i \chi_b^\dagger \left[ \left( A_\tau (0, \tau) \right)^b_c + v^I \left( \phi_I (0, \tau) \right)^b_c \right] \chi^c
\label{zsym}
\end{eqnarray}
Here $A_\mu$ and $\phi_I$ are bulk fields of SYM4 which are adjoints under SU($M$), $I = 1 \ldots 6$, 
$\mathcal{L}_{\rm SYM}$ is the Lagrangian of the bulk SYM4 CFT, 
and $v^I$ is a unit 6-vector
determining the specific choice of the quantum impurity. The similarity of Eq.~(\ref{zsym}) to Eqs.~(\ref{zphichi}), (\ref{znchi}),
and (\ref{zzchi}) should now be strikingly evident: in all cases we have an impurity localized fermions, and these are coupled to
the bulk CFT by a universal gauge-like coupling. While the present supersymmetric theory has no direct application
to condensed matter models, it has the advantage of being solvable by the AdS/CFT correspondence in the limit of $M \rightarrow \infty$.
Such a gravitational solution has been presented by Kachru {\em et al.} \cite{kky1,kky2}, who showed that the low energy
physics of the quantum impurity is associated with a AdS$_2$ geometry in the gravity 
theory (see also Refs~\cite{ads2a,ads2b,ads2c}). Further, the physical properties of the
model in Eq.~(\ref{zsym}) where found to be qualitatively identical to those listed above for Eq.~(\ref{zphichi}); in particular,
the AdS$_2$ solution also has a non-zero ground state impurity entropy.
Thus we may conclude \cite{ssffl} that there is an intimate connection between the quantum impurity models considered
in this section, and quantum gravity on AdS$_2$.

\subsection{Large $N$ solution}
\label{largen}

This subsection will illustrate the above general concepts on quantum impurities by an
explicit solution in a simple limiting case. We will look at a large $N$ limit in which a fermionic field is a vector
of $N$ components. It is also possible to set up a large $N$ limit which has the character of a matrix-large $N$ \cite{rs0},
but that is not easily solvable and will not be considered here.

We begin with the theory in Eq.~(\ref{zphichi}), and assume that the bulk field $\varphi^a$ has Gaussian correlations.
This neglects bulk interactions which are ultimately necessary for an accurate description of the critical properties; 
however, this omission will not be crucial for the calculation discussed below. Ultimately, the justification of non-Gaussian bulk
correlations relies on the large dimension or long-range limit of the models to be discussed in Section~\ref{sec:lattice} (note
that the bulk Gaussian approximation was not made in the theoretical studies noted above \cite{eps1,eps2,eps3,dc1,dc2,dc3,dc4}).
Anticipating Section~\ref{sec:lattice}, we assume that the correlation of $\varphi^a$ for the 
impurity physics are
fully characterized by the 2-point correlation
\begin{equation}
\left \langle \varphi^a (0, \tau) \varphi^b (0, \tau') \right\rangle = \delta^{ab} D (\tau - \tau')
\label{phiD}
\end{equation}
Integrating out $\varphi^a$ from Eq.~(\ref{zphichi}), we obtain a `local' partition function which involves a functional integral
over fields that depend only upon $\tau$
\begin{eqnarray}
\mathcal{Z} &=& \int \mathcal{D} \chi^\alpha (\tau) \mathcal{D} \lambda (\tau) \exp \left( 
- \int d \tau \left( \chi_\alpha^\dagger \frac{\partial \chi^\alpha}{\partial \tau} + i \lambda ( \chi^\dagger_\alpha \chi^\alpha - N/2) \right) \right. \nonumber \\
&~&~~\left. + \frac{2J^2}{N} \int d \tau d \tau' D(\tau - \tau') \chi_\alpha^\dagger (\tau) \chi_\beta^\dagger (\tau') \chi^\beta (\tau) \chi^\alpha (\tau') \right)
\label{zchi2}
\end{eqnarray}
where the indices $\alpha, \beta = 1 \ldots N=2$. However, we have written the partition function in a manner
so that it can be used for general $N$, and the limit $N \rightarrow \infty$ is well-defined. Indeed, an explicit solution
can be obtained in the large $N$ limit \cite{sy}, as we now describe.

\begin{figure}
\centering
 \includegraphics[width=4in]{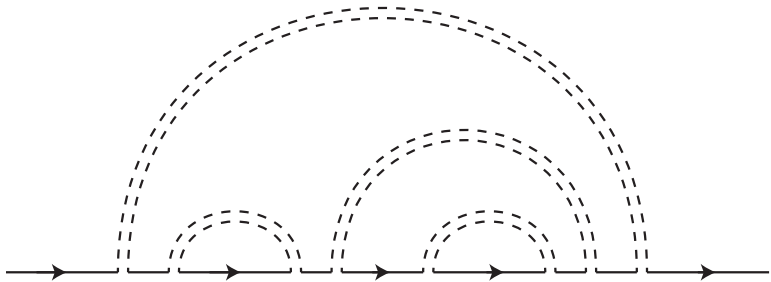}
 \caption{The rainbow graphs. The full line is the fermion, and double-dashed line is the interaction $D(\tau)$.
 Each line, full or dashed, carries a SU($N$) index $\alpha$.}
\label{fig:rainbow}
\end{figure}

An examination of the Feynman graph expansion shows that the limit of large $N$ is dominated by \cite{cox}
the `rainbow' (or `non-crossing') graphs for the fermion Green's function: see Fig.~\ref{fig:rainbow}. 
The summation of these graphs
can be written analytically in terms of the following self-consistency conditions on the fermion self energy.
As usual, we define the fermion Green's function G by
\begin{equation}
G(\tau) \delta_\beta^\alpha = - \left\langle \mathcal{T} \chi^\alpha (\tau) \chi^\dagger_\beta (0) \right\rangle
\quad , \quad G (\omega_n) = \int_0^{1/T} d\tau e^{i \omega_n \tau} G(\tau)
\label{g1}
\end{equation}
where $\mathcal{T}$ is imaginary-time ordering, and $\omega_n$ is a Matsubara frequency. This Green's function is expressed in terms
of the self energy by
\begin{equation}
G ( \omega_n) = \frac{1}{i \omega_n - \overline{\lambda} - \Sigma (\omega_n) }
\label{g2}
\end{equation}
where $\overline{\lambda}$ is the saddle-point value of $i \lambda$. Then, in the large $N$ limit, it is not
difficult to show that the fluctuations of $\lambda$ about its saddle point can be neglected, and 
the self energy is given by
\begin{equation}
\Sigma (\tau) = 4 J^2 D (\tau) G (\tau).
\label{g3}
\end{equation}
Finally, the fermion number constraint $n_\chi = N/2$, is equivalent to
\begin{equation}
G (\tau \rightarrow 0 ) = -\frac{\mbox{sgn} (\tau)}{2}.
\label{g4}
\end{equation}
   
We are now faced with the mathematical problem of solving the Eqs~(\ref{g2},\ref{g3},\ref{g4}) for the unknown
functions $G(\tau)$ and $\Sigma (\tau)$, and the value of $\overline{\lambda}$. While this is not difficult to do
numerically \cite{sy} for a general $D(\tau)$, we present here an analytic solution in the limit of low energies for
the case of critical correlations in $D(\tau)$. When the bulk theory is a CFT, we expect a power-law decay
$D (\tau ) \sim \tau^{-\gamma}$, with $\gamma$ a critical exponent. We generalize this to $T>0$, with the
`conformal' form
\begin{equation} 
D (\tau) = A \left| \frac{ \pi T}{\sin (\pi T \tau)} \right|^{\gamma} \quad , \quad -1/T < \tau < 1/T
\label{dt}
\end{equation}
where $A$ is some real constant. This is the general form of a $T>0$ correlator at $x=0$ for a CFT2, and holds also
for $\mathcal{L}_\varphi$ in the upper-critical dimension $D=4$ which is the only case where
the present Gaussian approximation for $\varphi^a$ correlations is appropriate.
Note that Eq.~(\ref{dt}) is supposed to be valid at energies well below the ultraviolet cutoff $\sim J$;
in other words, for $1/|\tau|, T \ll J$.

We will now show that the solution of Eqs~(\ref{g2},\ref{g3},\ref{g4}) has the following form 
at long times \cite{sy,pgks}
\begin{equation}
G (\tau) = B \, \mbox{sgn} (\tau) \left| \frac{ \pi T}{\sin (\pi T \tau)} \right|^{\rho} \quad , \quad -1/T < \tau < 1/T
\label{gt}
\end{equation}
and determine the exact values of $B$ and the exponent $\rho$. Again Eq.~(\ref{gt}) holds only for low energies
with $1/|\tau|, T \ll J$. We have used the particle-hole symmetric nature of the constraint in Eq.~(\ref{g4}) to conclude
that $G$ should be an odd function of $\tau$. Also, note that we are not concerned that Eq.~(\ref{gt}) does not
obey Eq.~(\ref{g4}) as $\tau \rightarrow 0$, because Eq.~(\ref{gt}) does not apply in this limit.

We can now perform the Fourier transform of Eqs.~(\ref{g3},\ref{dt},\ref{gt}) to obtain the low frequency behavior
of the Green's function and the self energy:
\begin{eqnarray}
G (\omega_n) &=& \left[ i  B  \Pi (\rho) \right]  \frac{ T^{\rho-1} \,
 \Gamma \left( \displaystyle  \frac{\rho}{2}  + \frac{\omega_n}{2 \pi T} \right)}{
 \Gamma \left( \displaystyle 1 - \frac{\rho}{2} + \frac{\omega_n}{2 \pi T} \right)  } 
 \nonumber \\
\Sigma_{\rm sing} (\omega_n ) &=& \left[ i   4 J^2 A B  \Pi (\rho+\gamma) \right]  \frac{  T^{\rho+\gamma-1} \,
 \Gamma \left( \displaystyle  \frac{\rho+\gamma}{2}  + \frac{\omega_n}{2 \pi T} \right)}{
 \Gamma \left( \displaystyle 1 - \frac{\rho+\gamma}{2} + \frac{\omega_n}{2 \pi T} \right)  } 
 \label{gs1}
\end{eqnarray}
with 
\begin{equation}
\Pi (s) \equiv \pi^{s-1 } 2^s  \cos \left(\displaystyle  \frac{\pi s}{2} \right) \Gamma (1-s).
\end{equation}
We have noted that the contribution to $\Sigma$ in Eq.~(\ref{gs1}) is only the singular low frequency term,
and an additional cutoff-dependent constant has been omitted.

It now remains to determine if the proposed solution in Eq.~(\ref{gs1}) obeys the Dyson equation in Eq.~(\ref{g2}).
We need only obtain agreement in the low frequency limit, where we find that after canceling $\overline{\lambda}$ 
with the regular part of the self-energy, the Dyson equation reduces for the singular contributions simply to \cite{sy}
\begin{equation}
G (\omega_n) \Sigma_{\rm sing} (\omega_n ) = -1.
\label{gs2}
\end{equation}
Note that we have assumed that the bare $i \omega_n$ frequency dependence in Eq.~(\ref{g2}) is sub-dominant
to the singular contribution from the self energy in the low frequency limit; this requires $\gamma < 2$.
Remarkably, we find that the frequency dependent expressions in Eq.~(\ref{gs1}) can indeed satisfy the constraint
in Eq.~(\ref{gs2}) for {\em all\/} $\omega_n$ provided we choose the exponent \cite{sy,pgks}
\begin{equation}
\rho = 1 - \frac{\gamma}{2}, 
\end{equation}
and the prefactor
\begin{equation}
B = \left[ 4 J^2 A \Pi (1 - \gamma/2) \Pi (1 + \gamma/2) \right]^{-1/2}.
\end{equation}
Thus we have the surprising conclusion that the innocuous-looking Eqs~(\ref{g2},\ref{g3},\ref{g4}) which solve 
the partition function in Eq.~(\ref{zchi2}) in the large-$N$ limit
have a solution which has a conformally-invariant structure at low energies \cite{pgks,pgs}. 
We then observe that the supersymmetric model in Eq.~(\ref{zsym}) also has correlators which have a conformal
structure, inherited from the AdS$_2$ geometry \cite{Faulkner,Iqbal}. This is then further evidence for the striking connection between the
quantum impurity models of this section and the theories of gravity on AdS$_2$.

It is also useful to collect results for the impurity spin correlation function
\begin{equation}
C(\tau) = \left \langle \hat{S}^a (\tau)  \hat{S}^a (0) \right \rangle
\label{c1}
\end{equation}
from the present large $N$ solution. Using the SU(2) relation $\hat{S}^a = (1/2) \chi_\alpha^\dagger (\sigma^a )^{\alpha}_{\beta} \chi^\beta$,
and evaluating Eq.~(\ref{c1}) in the large $N$ limit, we find $C(\tau) = - (N^2/2) G(\tau) G(-\tau)$, and so
\begin{equation}
C (\tau) = \frac{B^2 N^2}{2} \left| \frac{ \pi T}{\sin (\pi T \tau)} \right|^{h} 
\label{ct}
\end{equation}
where the exponent \cite{sy,anirvan}
\begin{equation}
h = 2 - \gamma.
\label{h1}
\end{equation}
Of course, in the large $N$ limit we have $ h = 2 \rho$, but we have introduced an independent exponent $h$ for $C(\tau)$
because we expect that at higher orders in $1/N$ we have $h \neq 2 \rho$.
On the other hand, it has been argued \cite{eps2} that the exponent relationship in Eq.~(\ref{h1}) is {\em exact\/}, 
and holds to all orders in the $1/N$ expansion for the theory in Eq.~(\ref{zchi2}).
This exponent relationship is a consequence of the fact that the Gaussian field $\varphi^a$ and the spin operator
$\hat{S}^a$ are conjugate operators in the second term of $\mathcal{L}_{\rm imp}$ in Eq.~(\ref{zphi}), and 
the co-efficient of this term reaches a fixed-point value in the RG.

Finally, this large $N$ computation can also be used to compute the impurity entropy.
This requires a somewhat more involved computation \cite{pgks,pgs}, and will not be presented here.

\section{From a quantum impurity to lattice models}
\label{sec:lattice}

The quantum impurity models discussed in Section~\ref{sec:qimp} now appear to be very well understood: the results
of different expansion methods are consistent with each other, and with a variety of numerical studies. And for the supersymmetric
impurity models, the AdS/CFT method yields similar results using the geometry of AdS$_2$.

In this section, we move beyond impurity models to a variety of lattice models. In condensed matter studies, 
this is done by using various expansion methods or 
physical arguments to motivate a specific mean-field decoupling of the quantum lattice model to a preferred `impurity' spin coupled
to a bulk `environment'. The resulting mean-field theory then has a structure similar to the models considered in Section~\ref{sec:qimp}.
However, now the `environment' is built out of the same degrees of freedom that yielded the `impurity' spin. 
This fact leads to an additional self-consistency condition that supplements the solution of the impurity model; it is
their combination which then yields the mean-field predictions for the quantum lattice model. 

As we will see below, the resulting mean-field theory of the lattice model has strong similarities to the 
classical gravity theories of AdS$_{D+1}$ at non-zero $\mu$ which were outlined in Section~\ref{sec:intro}. 
In the latter theories, the geometry factorizes to AdS$_2 \times R^{D-1}$ at low energies; it is this factorization
which will be seen to be related to the `impurity' + `environment' factorization of the condensed matter mean-field theories.
A notable fact is that the factorization is motivated in the gravity theory from a very different reasoning from
that in the condensed matter model. Thus the appearance of similar result in two very different approximations is
quite surprising, and indicates a robustness of the theory that should be well worth understanding better.

Returning to the condensed matter perspective, let us describe the mapping from lattice models to self-consistent quantum impurity
models. Such a mapping has been carried out for a variety of models \cite{sy,rahul,qmsi,chitra,si1,bgg,si2},
which are all versions of the Kondo lattice Hamiltonian
\begin{equation}
H = H_J + \sum_k \epsilon_k c^{\dagger}_{k \alpha} c_{k}^{\alpha} + \frac{J_K}{2} \sum_i \hat{S}^a_i c^{\dagger}_{i \alpha} 
(\sigma^a)^{\alpha}_{\beta} c_{i}^{\beta}.
\end{equation}
Here $H_J$ is a quantum spin model just as in Eq.~(\ref{eq:ssHJ}). To these localized spins, we have
added mobile conduction electrons: $c_{k}^{\alpha}$ is the Fourier transform of the electron operator $c_i^\alpha$ 
on site $i$ and $\epsilon_k$ is the electron dispersion. Finally.
$J_K$ is the Kondo exchange coupling between the conduction electrons and the spins.

The most direct mean-field analyses of $H$ appear in 
lattice models with random infinite-range exchange interactions \cite{sy,bgg}: the $J_{ij}$ being independent
Gaussian random variables with zero mean. Note that the disorder is `quenched' {\em i.e.\/} each $J_{ij}$ is independent
of time, but is chosen at random from a Gaussian distribution.  However, similar mean field equations also arise
in the large spatial dimension limit of non-random Kondo lattice models \cite{qmsi,chitra,si1,bgg,si2}.

Such mean field models yield solutions corresponding to the two classes of non-magnetic metallic states
expected in Kondo lattice models \cite{ffl1,ffl2,vojtarev}:
\begin{itemize}
\item
A Fermi liquid (FL) with a `large Fermi surface', which can be viewed as arising from the RG flow to large $J_K$. Here the electrons
forming the $\hat{S}^a_i$ spins, along with the $c_{i}^\alpha$ electrons, become part of the Luttinger count which determines the volume enclosed by the Fermi surface. 
\item
A fractionalized Fermi liquid (FFL or FL*) with a `small Fermi surface', in which the effects of $J_K$ can be accounted
for perturbatively. Here the $\hat{S}^a$ spins form a spin liquid, while the conduction electrons form a Fermi surface whole volume
counts only the density of the $c_{i}^\alpha$ conduction electrons.
\end{itemize}

Let us describe the mean-field structure of the FL* phase so obtained \cite{sy,bgg}.
It was found that correlations of the spin liquid sector of this phase are described by a quantum
impurity theory which is identical to Eq.~(\ref{zchi2}). However, this theory now has an additional self-consistency condition
that the `environment' spins $\varphi^a$ are the same as the impurity spin $\hat{S}^a$: thus the two-point correlator
of $\varphi^a$ which appears in Eqs~(\ref{phiD}) and (\ref{zchi2}) should be proportional to the two-point correlator
of $\hat{S}^a$ in Eq.~(\ref{c1}): {\em i.e.\/}
\begin{equation}
D(\tau) \propto C (\tau).
\end{equation}
It was further shown that a solution of this self-consistency relation is only possible if the spectrum is gapless and 
has a power-law form. Then from Eqs.~(\ref{dt}) and (\ref{ct}) we have the exponent relation
\begin{equation}
h = \gamma.
\end{equation}
Combining this with the exact relation in Eq.~(\ref{h1}) for the quantum impurity model, we obtain \cite{sy}
\begin{equation}
h=1.
\label{heq1}
\end{equation}
This is the value that corresponds to `marginal Fermi liquid' behavior \cite{varma}, as we will see shortly.
The same value of $h$ is obtained in large dimension solution of non-random lattice models \cite{si1,si2}.
The gravity approach does not fix the value of such exponents: they are related to the mass of fermions
in AdS$_{D+1}$ and these are free parameters in present analyses.

We have now described a mean-field state, which applies to both random and non-random models,
which is a critical spin liquid. It can be viewed as a large dimension analog of the `Spin Bose Metal' \cite{mpaf1,mpaf2},
in that it can be written as a theory of bosons at non-zero density which do not Bose condense, but form
a gapless liquid \cite{sy}.
This mean-field state has a non-zero ground state entropy density, which descends 
from the entropy of the quantum impurity problem.
It is this critical spin liquid which has been proposed to realize the low energy physics of AdS$_{D+1}$ in a 
non-zero chemical potential $\mu$.

The main claim of Ref.~\cite{ssffl} was that the theory of the holographic metal realizes the FL* phase, in a situation in which
the $\hat{S}^a_i$ spins are in the critical spin liquid state described in Section~\ref{largen}.  
The evidence for this claim so far is the gapless conformal form of the spin and fermion correlations in 
Section~\ref{largen}, the connection with AdS$_2$ of the impurity models in Section~\ref{sec:qimp},
and the non-zero ground state entropy density. Additional evidence comes from the 
self-energy of the conduction electrons, $c_i^\alpha$. 
We can compute the conduction electron self-energy $\Sigma_c (\omega_n)$
in the FL* phase by perturbation theory in $J_K$; at second order in $J_K$ we have the contribution \cite{bgg}
\begin{eqnarray}
\Sigma_c (\omega_n) &\propto& J_K^2 T \sum_{\epsilon_n} \sum_k \frac{1}{i (\omega_n + \epsilon_n) - \epsilon_k} \, C (\epsilon_n) 
\nonumber \\
&\propto & - i \pi N_0 J_K^2 T \sum_{\epsilon_n} \mbox{sgn} (\omega_n + \epsilon_n) C (\epsilon_n) \nonumber \\
&=& J_K^2 T^h \Psi ( \omega_n / T)
\label{selfc}
\end{eqnarray}
where $\Psi (\omega_n / T)$ is a scaling function, and $N_0$ is the density of conduction electron states at the Fermi level.
Provided $h < 2$, this is a non-Fermi liquid form of the electron self energy.
It is also the same result as that obtained for the holographic metal in Refs.~\cite{Faulkner,mfl,physics}. A similar analysis can be done
by the present methods of the transport properties \cite{bgg}, and again agreement is obtained with the holographic
results \cite{Faulkner,mfl,physics}.

For the exponent $h=1$ obtained \cite{sy} by the self-consistency requirement on lattice models in Eq.~(\ref{heq1}), 
the self energy in Eq.~(\ref{selfc})
has the marginal Fermi liquid form \cite{varma}.

\section{Conclusion}

In conclusion, we note that theory of the FL* phase reviewed here \cite{sy,bgg}, and obtained by the mapping to quantum impurity models,
is an attractive candidate for describing strange metal phases. At the semi-phenomenological level, it does provide
a satisfactory description of experimental observations. It is indeed quite remarkable and surprising that a similar
theory has now appeared from the very different starting point of the AdS/CFT correspondence. 

However, the
quantum-impurity description of the FL* phase is not believed to be complete \cite{ffl1,ffl2}. The spin liquid constituent of this phase 
should have emergent gauge
excitations (like the $A_\mu$ gauge field in Eq.~(\ref{zz})), and these are surely essential for a complete description of spatial correlations. So it would be interesting
to find the appropriate gauge fields in the holographic theory. In this context, the recent work of Nickel and 
Son is notable \cite{sonrecent}, as they argue
that theory of the holographic metal may indeed be missing such emergent gauge fields.

We also note the interesting recent work of Kachru {\em et al.} \cite{kky2} showing a transition from a FL* phase to a FL phase using string theory.

\subsection*{Acknowledgements}
I am grateful to Sean Hartnoll and Shamit Kachru for very helpful discussions.
This research was supported by the National Science Foundation under grant DMR-0757145 and by a MURI grant from AFOSR.

\end{document}